\documentclass[conference,10pt]{IEEEtran}
\IEEEoverridecommandlockouts

\usepackage{epsfig,rotating,setspace,latexsym,amsmath,epsf,amssymb,amsfonts,bm,theorem,cite,algorithm,graphicx,epsf,authblk,epstopdf,color,algpseudocode,bbm}

\newtheorem{theorem}{Theorem}
\newtheorem{lemma}{Lemma}
\newenvironment{Proof}[1]{\medskip\par\noindent{\bf Proof:\,}\,#1}{{\mbox{\,$\blacksquare$}\par}}

\allowdisplaybreaks

\begin{document}

\title{On Timely Channel Coding with Hybrid ARQ\thanks{This work was supported in part by the U.S. National Science Foundation under Grants CCF-0939370 and CCF-1513915.}\vspace{-.1in}}


\author[1]{Ahmed Arafa}
\author[2]{Karim Banawan}
\author[3]{Karim G. Seddik}
\author[1]{H. Vincent Poor\vspace{-.1in}}
\affil[1]{\normalsize Electrical Engineering Department, Princeton University, USA}
\affil[2]{\normalsize Department of Electrical Engineering, Alexandria University, Egypt}
\affil[3]{\normalsize Electronics and Communications Engineering Department, American University in Cairo, Egypt}

\maketitle

\begin{abstract}
A status updating communication system is examined, in which a transmitter communicates with a receiver over a noisy channel. The goal is to realize timely delivery of fresh data over time, which is assessed by an {\it age-of-information} (AoI) metric. Channel coding is used to combat the channel errors, and feedback is sent to acknowledge updates' reception. In case decoding is unsuccessful, a hybrid ARQ protocol is employed, in which incremental redundancy (IR) bits are transmitted to enhance the decoding ability. This continues for some amount of time in case decoding remains unsuccessful, after which a new (fresh) status update is transmitted instead. In case decoding is successful, the transmitter has the option to idly {\it wait} for a certain amount of time before sending a new update. A general problem is formulated that optimizes the codeword and IR lengths for each update, and the waiting times, such that the long term average AoI is minimized. {\it Stationary deterministic} policies are investigated, in which the codeword and IR lengths are fixed for each update, and the waiting time is a deterministic function of the AoI. The optimal waiting policy is then derived, and is shown to have a {\it threshold} structure, in which the transmitter sends a new update only if the AoI grows above a certain threshold that is a function of the codeword and IR lengths. Choosing the codeword and IR lengths is discussed in the context of binary symmetric channels.
\end{abstract}

\section{Introduction}

Real-time status updating systems require timely information delivery of fresh data to interested destinations. A suitable metric to assess such timeliness/freshness is the {\it age-of-information} (AoI) metric, defined as the time elapsed since the latest status update has reached the destination. Considering AoI as a performance metric for timely information delivery has been considered under various settings in the recent literature, covering topics in queuing, scheduling and optimization to assess and improve data freshness, see, e.g., \cite{yates_age_1, ephremides_age_random, chen-age-error, ephremides_age_non_linear, shroff_age_multi_hop, sun-age-mdp, talak-age-interference, zhou-age-iot, jing-age-online, arafa-age-online-finite, elif-age-online-threshold}. Of particular relationship to this work are those pertaining to coding for AoI improvement \cite{najm-age-mg11-harq, sac-age-mg1-harq, simeone-age-finite-code, najm-age-erasure-coding, parag-age-coding, yates-age-erase-code, ceran-age-harq, baknina-age-coding, feng-age-rateless-codes}. 

The works in \cite{najm-age-mg11-harq, sac-age-mg1-harq, simeone-age-finite-code, najm-age-erasure-coding} have the common feature that updates are externally arriving. Specifically, \cite{najm-age-mg11-harq} analyzes AoI in an $M/G/1/1$ queue, in which updates are sent through an erasure channel using different hybrid ARQ (HARQ) protocols, with and without preemption. \cite{sac-age-mg1-harq} considers an $M/G/1$ queue, under a first-come first-serve (FCFS) discipline, and analyzes both average and peak AoI. \cite{simeone-age-finite-code} considers sending updates over an additive white Gaussian noise channel with ARQ, with updates arriving according to a Bernoulli process and a FCFS discipline, leveraging finite blocklength information-theoretic results to characterize peak AoI and peak delay violation probabilities. \cite{najm-age-erasure-coding} follows an information-theoretic approach to analyze age-minimal coding design in erasure channels without feedback, for given update generation and channel usage rates, and source and channel alphabets.

On the other hand, status updates in \cite{parag-age-coding, yates-age-erase-code, ceran-age-harq, baknina-age-coding, feng-age-rateless-codes} can be generated {\it at will}. Reference \cite{parag-age-coding} analyzes the effect of linear block coding lengths on AoI in erasure channels, with and without HARQ. \cite{yates-age-erase-code} compares two coding techniques for erasure channels: infinite incremental redundancy (IIR), in which a rateless code is used to send an update until it is successfully decoded; and fixed redundancy, in which an update is encoded using a fixed codeword length. \cite{ceran-age-harq} studies AoI minimization in erasure channels under an average constraint on the number of transmissions, with ARQ and HARQ, formulated as a constrained Markov decision process (MDP). \cite{baknina-age-coding} considers a more specific energy harvesting constraint, along with both rateless and maximum distance separable (MDS) codes, and derives achievable AoI under best effort and save-and-transmit strategies. \cite{feng-age-rateless-codes} considers IIR HARQ in erasure channels, with the option of preempting the current update in service and switching to a new one, through an MDP framework.

In this paper, we consider a transmitter-receiver pair communicating through a noisy channel. The main goal is to keep the receiver informed about the status of some physical phenomenon over a long period of time, via sending time-stamped status updates. Updates are generated at will, and are encoded to combat the channel errors. Different from most related works that focus on erasure channels, we consider channels with general error models that depend on the coding lengths. Utilizing decoding status feedback, a HARQ protocol is employed, in which a number of incremental redundancy (IR) bits is sent in case decoding is unsuccessful. Different from \cite{parag-age-coding, yates-age-erase-code, ceran-age-harq, baknina-age-coding, feng-age-rateless-codes}, the transmitter is allowed to idly {\it wait} for a period of time following successful transmission, and then send a new update. We use an AoI metric to assess the timeliness of the received updates, where the goal is to design the codeword and IR lengths, in addition to the waiting times, such that the long term average AoI is minimized. We focus on stationary deterministic policies, in which the codeword and IR lengths are fixed, and the waiting time is a deterministic function of the instantaneous AoI. We show that the optimal waiting policy in this case has a {\it threshold} structure, in which a new update is sent only if the AoI grows above a certain threshold that is a function of the codeword and IR lengths. We analytically derive the optimal threshold in {\it closed-form}, and discuss some examples showing how to choose the best codeword and IR lengths for a binary symmetric channel.

\section{System Model and Problem Formulation}

\begin{figure}[t]
\center
\includegraphics[scale=.7]{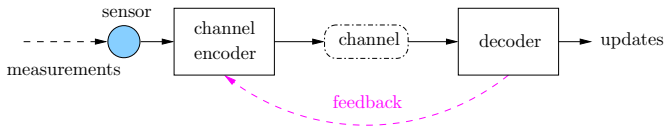}
\caption{Measurements are acquired by the sensor, encoded, sent through the channel and then decoded at the receiver to produce updates. Feedback indicates successful/failed decoding attempts.}
\label{fig_sys_mod}
\vspace{-.1in}
\end{figure}

The transmitter that we consider mainly consists of a sensor and a channel encoder, and the receiver is mainly a channel decoder, see Fig.~\ref{fig_sys_mod}. Status updates are generated at will through collecting measurements of the physical phenomenon by the sensor, and are conveyed to the receiver through a noisy communication channel. Updates are basically data packets that contain a time stamp indicating when their corresponding measurements were acquired. We capture the timeliness/freshness of data packets at the receiver using an AoI metric, which is defined at time $t$ as
\begin{align}
a(t)=t-u(t),
\end{align}
where $u(t)$ denotes the time stamp of the latest data packet that has been successfully received before time $t$. Operationally, for the receiver to be informed about the process, the AoI (or merely {\it age}) needs to be kept small.

Channel coding with HARQ is employed in order to combat the noisy communication channel. This is illustrated as follows. The $i$th raw measurement from the sensor is first converted into an update data packet of length $\ell_i$ bits. This packet gets mapped into a codeword of length $n_i$ bits, to be sent through the channel. Different from the FR scheme in \cite{yates-age-erase-code}, the receiver attempts decoding after receiving the whole codeword. The receiver then sends immediate feedback ACK/NACK messages to the transmitter following successful/failed decoding attempts. The HARQ protocol is such that the $i$th update can have at most $r_i+1$ decoding attempts at the receiver. Specifically, if the first decoding attempt fails (after receiving the original $n_i$-bit codeword), the receiver sends a NACK, and the transmitter responds by sending $m_{i,1}$ IR bits. The receiver then combines the originally received $n_i$ bits and the newly received $m_{i,1}$ bits to perform a second decoding attempt. If it also fails, the receiver sends another NACK, and the transmitter responds again by sending $m_{i,2}$ IR bits. This continues until either the packet is successfully decoded and the receiver sends an ACK, or the maximum number of $r_i+1$ decoding attempts is reached. Therefore, in total there can be at most $r_i$ IR transmissions for the $i$th update, with the maximum total number of bits used to transmit the $i$th update in this case equal to $n_i+\sum_{j=1}^{r_i}m_{i,j}$.


Now if all the $r_i+1$ decoding attempts fail, the $i$th packet is discarded, and the whole process is repeated with a new {\it fresher} $(i+1)$th measurement.\footnote{We note that the values of $n_i$'s, $r_i$'s and $m_{i,j}$'s, are all pre-determined before communication and shared with both the transmitter and the receiver.} On the other hand, if an ACK is received for the $i$th update (following any of the decoding attempts), the transmitter may not send a new packet right away; it may idly wait for some time before doing so. In general, this waiting period may differ from one packet to the other, and may also depend on the number of decoding attempts involved for each packet.

The communication channel is assumed to be memoryless, and is such that the {\it success} probability of decoding an update {\it increases} with each decoding attempt, and also increases with the number of bits involved in its transmission (the sum of the original codeword's length and the IR bits). We denote the success probability of decoding the $i$th update in its $j$th attempt by $q_{i,j}$, $j=1,\dots,r_i+1$. We assume that the time needed to transmit one bit through the channel is normalized, in the sense that sending, e.g., $n_i$ bits consumes $n_i$ time units.

\begin{figure}[t]
\center
\includegraphics[scale=.8]{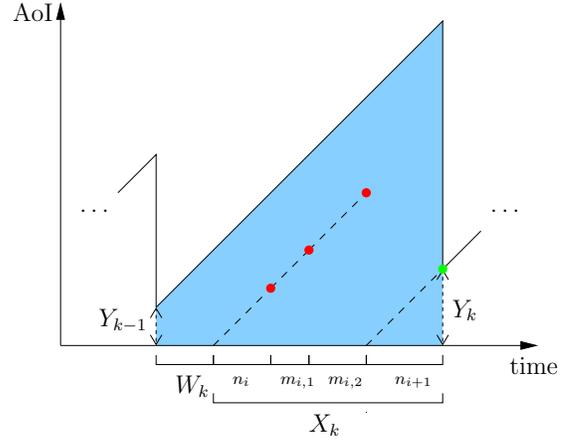}
\caption{An example showing how the AoI may evolve during the $k$th epoch. The epoch involves two update transmissions: the $i$th and the $(i+1)$th updates. Red (reps. green) dots represent failed (resp. successful) decoding attempts.}
\label{fig_age_xmpl}
\vspace{-.2in}
\end{figure}

We denote by an {\it epoch} the time elapsed in between two successful receptions of update data packets. We now introduce some notation regarding the $k$th epoch. The epoch starts with age equal to $Y_{k-1}$, followed by a potential waiting period of $W_k$ time units, which is then followed by the {\it channel busy period} $X_k$, whose value depends on the number of transmissions needed before decoding is successful. The epoch ends with age $Y_k$. Let us denote by $Q_k$ the area under the AoI evolution curve during the $k$th epoch, and by $L_k$ the $k$th epoch length. In Fig.~\ref{fig_age_xmpl}, we show an example of how the age may evolve during the $k$th epoch.

For a given coding scheme, what is designed for each update are the original codewords' lengths $\{n_i\}$, the number of IR transmissions $\{r_i\}$ and the lengths of the IR bits $\{m_{i,j}\}$, along with the waiting times $\{W_k\}$ following each successful transmission. We denote all these by a policy $\pi$, and the set of all possible policies by $\Pi$. Our goal is to design a policy such that the long term average AoI is minimized. That is, to solve the following problem:
\begin{align} \label{opt_main}
\min_{\pi\in\Pi}\quad\limsup_{l\rightarrow\infty}\frac{\sum_{k=1}^l\mathbb{E}\left[Q_k\right]}{\sum_{k=1}^l\mathbb{E}\left[L_k\right]},
\end{align}
where $\mathbb{E}\left[\cdot\right]$ denotes expectation.

\section{Stationary Deterministic Policies}

Observe that in the optimal policy of problem (\ref{opt_main}), the choices of the $i$th update parameters, $n_i$, $r_i$ and $m_{i,j}$, $j=1,\dots,r_i$, may depend on the history of events prior to the $i$th update transmission, e.g., how many successful/failed attempts occurred. Similarly, the choice of the waiting time in the $k$th epoch, $W_k$, may also depend on the history of events in previous epochs. To alleviate this hurdle and make problem (\ref{opt_main}) tractable, we focus on a class of {\it stationary deterministic} policies, in which the choices of parameters are {\it fixed} for each update, and the waiting policy is a deterministic function $w(\cdot)$ of the epoch's starting AoI, i.e.,
\begin{align}
W_k\triangleq w\left(Y_{k-1}\right), \quad \forall k.
\end{align}
This induces stationary distributions $\{Q_k\}$ $\sim Q$ and $\{L_k\}$ $\sim L$ over all epochs. Considering such class of policies is motivated by the fact that the channel is memoryless, and also by its optimality in similar settings considered in the literature, e.g., \cite{sun-age-mdp}, and the optimality of renewal policies in \cite{jing-age-online, arafa-age-online-finite}. We also focus on the case in which each packet has a maximum of two decoding attempts, i.e., $r_i=1$, $\forall i$. We note that our results can be readily extended for more than two decoding attempts per single packet, albeit more involved computations. We choose to work with only two attempts in order to better convey the main ideas of this paper.

It is important to note that while \cite{sun-age-mdp} determines optimality conditions on the waiting policy, its setting is different from ours. Specifically, in \cite{sun-age-mdp} there can only be one packet transmission during an epoch, and it stays in service (for some random time) until it reaches the destination. While in our setting, there can be multiple packet transmission {\it attempts} during one epoch, and the AoI at the end of the epoch does {\it not} necessarily correspond to the service time of the first packet transmission.

Focusing on stationary deterministic policies, we now drop all subscripts, and (re-)define: $\ell$ as the update data packet's length, $n$ as the codeword length, $m$ as the number of IR bits, $q_1$ (resp. $q_2$) as the success probability of the first (resp. second) decoding attempt and $X$ as the channel busy period. Each epoch now starts with AoI $Y$ that is either equal to $n$ or $n+m$. The waiting policy reduces to
\begin{align}
w\left(Y\right)=\begin{cases}w_1,\quad&\text{if }Y=n\\
w_2,\quad&\text{if }Y=n+m\end{cases}.
\end{align}
That is, $w_1$ (resp. $w_2$) is the waiting time following successful decoding of a packet from the first (resp. second) attempt. Problem (\ref{opt_main}) now reduces to an optimization over a single epoch as follows:
\begin{align} \label{opt_sd}
\min_{n,m,w_1,w_2}\quad&\frac{\mathbb{E}\left[Q\right]}{\mathbb{E}\left[L\right]} \nonumber \\
\mbox{s.t.}\hspace{.15in}\quad&n,m\in\mathbb{Z}_+ \nonumber \\
&w_1,w_2\geq0,
\end{align}
where $\mathbb{Z}_+$ is the set of non-negative integers. We focus on problem (\ref{opt_sd}) in the remainder of this paper.

\section{Optimal Waiting Policies:\\Threshold Structure}

In this section, we derive the optimal waiting policy $w_1^*$ and $w_2^*$ that solves problem (\ref{opt_sd}) for fixed $n$ and $m$ and show that it has a threshold structure. We first observe from Fig.~\ref{fig_age_xmpl} that
\begin{align} \label{eq_exp_L}
\mathbb{E}\left[L\right]=&\mathbb{E}\left[X\right]+\mathbb{E}\left[w\left(Y\right)\right], \\
\mathbb{E}\left[Q\right]=&\mathbb{E}\left[Y\left(w\left(Y\right)+X\right)\right]+\frac{1}{2}\mathbb{E}\left[\left(X+w\left(Y\right)\right)^2\right] \nonumber \\
=&\mathbb{E}\left[Yw\left(Y\right)\right]+\mathbb{E}\left[Y\right]\mathbb{E}\left[X\right]+\frac{1}{2}\mathbb{E}\left[X^2\right] \nonumber \\
&+\mathbb{E}\left[X\right]\mathbb{E}\left[w\left(Y\right)\right]+\frac{1}{2}\mathbb{E}\left[w\left(Y\right)^2\right]. \label{eq_exp_Q}
\end{align}

Toward characterizing $\mathbb{E}\left[L\right]$ and $\mathbb{E}\left[Q\right]$ in terms of $n$, $m$, $w_1$ and $w_2$, note that the channel busy period, $X$, has the following distribution for a given $j\geq1$:
\begin{align} \label{eq_dist_X}
\mathbb{P}\left(X=jn+(j-1)m\right)=&(1-q_1)^{j-1}(1-q_2)^{j-1}q_1, \\
\mathbb{P}\left(X=jn+jm\right)=&(1-q_1)^j(1-q_2)^{j-1}q_2.
\end{align}
Various quantities can now be computed using (\ref{eq_dist_X}). For instance, since the distributions are all stationary, the starting AoI of the epoch, $Y$, has the following distribution:
\begin{align}
\mathbb{P}\left(Y=n\right)=\sum_{j=1}^\infty\mathbb{P}\left(X=jn+(j-1)m\right)=\frac{q_1}{q_1+q_2-q_1q_2},
\end{align}
along with $\mathbb{P}\left(Y=n+m\right)=1-\mathbb{P}\left(Y=n\right)$. Using this, one can directly get
\begin{align}
\mathbb{E}\left[Y\right]=&n+m\frac{(1-q_1)q_2}{q_1+q_2-q_1q_2}, \label{eq_exp_Y} \\
\mathbb{E}\left[w\left(Y\right)\right]=&\frac{q_1w_1+(1-q_1)q_2w_2}{q_1+q_2-q_1q_2}, \\
\mathbb{E}\left[w^2\left(Y\right)\right]=&\frac{q_1w_1^2+(1-q_1)q_2w_2^2}{q_1+q_2-q_1q_2}, \\
\mathbb{E}\left[Yw\left(Y\right)\right]=&\frac{q_1w_1n+(1-q_1)q_2w_2(n+m)}{q_1+q_2-q_1q_2}.
\end{align}
Finally, after some involved algebraic manipulations,
\begin{align}
\mathbb{E}\left[X\right]=&\frac{n+m(1-q_1)}{q_1+q_2-q_1q_2}, \label{eq_exp_X} \\
\mathbb{E}\left[X^2\right]=&\frac{\left(n+m\right)^2\left(2-q_1-q_2+q_1q_2\right)-2m\left(n+m\right)q_1}{\left(q_1+q_2-q_1q_2\right)^2} \nonumber \\
&+\frac{m^2q_1}{q_1+q_2-q_1q_2}. \label{eq_exp_X2}
\end{align}

Substituting (\ref{eq_exp_Y})-(\ref{eq_exp_X2}) in (\ref{eq_exp_L}) and (\ref{eq_exp_Q}), we can now fully characterize the objective function of problem (\ref{opt_sd}) in terms of $w_1$ and $w_2$ for fixed $n$ and $m$. To get a handle on such a fractional optimization problem, we follows Dinkelbach's approach \cite{dinkelbach-fractional-prog} and introduce the following auxiliary problem for some fixed parameter $\lambda\geq0$:
\begin{align} \label{opt_aux}
p(\lambda)\triangleq\min_{w_1,w_2}\quad&\mathbb{E}\left[Q\right]-\lambda\mathbb{E}\left[L\right] \nonumber \\
\mbox{s.t.}\quad&w_1,w_2\geq0.
\end{align}
One can show that: 1) $p(\lambda)$ is decreasing in $\lambda$; and 2) the optimal solution of problem (\ref{opt_sd}) (for fixed $n$ and $m$) is given by the unique $\lambda^*$ that solves $p(\lambda^*)=0$ \cite{dinkelbach-fractional-prog}. We now have the following lemma:

\begin{lemma} \label{thm_waiting}
The optimal solution of problem (\ref{opt_aux}) is given by
\begin{align}
w_1^*=&\left[\lambda-\mathbb{E}\left[X\right]-n\right]^+, \label{eq_w1_opt} \\
w_2^*=&\left[\lambda-\mathbb{E}\left[X\right]-n-m\right]^+, \label{eq_w2_opt}
\end{align}
where $\mathbb{E}\left[X\right]$ is given by (\ref{eq_exp_X}) and $\left[\cdot\right]^+\triangleq\max\left(\cdot,0\right)$.
\end{lemma}

\begin{Proof}
We show this by deriving the KKT optimality conditions for problem (\ref{opt_aux}) by the Lagrangian \cite{boyd}
\begin{align}
\mathcal{L}=\mathbb{E}\left[Q\right]-\lambda\mathbb{E}\left[L\right]-\eta_1w_1-\eta_2w_2,
\end{align}
where $\eta_1$ and $\eta_2$ are non-negative Lagrange multipliers. Expanding only the terms that depend on $w_1$ and $w_2$:
\begin{align}
\mathcal{L}=&\frac{q_1w_1n+(1-q_1)q_2w_2(n+m)}{q_1+q_2-q_1q_2}+\!\mathbb{E}\left[Y\right]\mathbb{E}\left[X\right]+\!\frac{1}{2}\mathbb{E}\left[X^2\right] \nonumber \\
&+\mathbb{E}\left[X\right]\frac{q_1w_1+(1-q_1)q_2w_2}{q_1+q_2-q_1q_2}+\!\frac{1}{2}\frac{q_1w_1^2+(1-q_1)q_2w_2^2}{q_1+q_2-q_1q_2} \nonumber \\
&-\lambda\mathbb{E}\left[X\right]-\lambda\frac{q_1w_1+(1-q_1)q_2w_2}{q_1+q_2-q_1q_2}-\eta_1w_2-\eta_2w_2.
\end{align}
Taking derivative of $\mathcal{L}$ with respect to $w_1$ and equating to $0$ we get that the optimal $w_1^*$ satisfies
\begin{align}
\frac{q_1n+\mathbb{E}\left[X\right]q_1+q_1w_1^*-\lambda q_1}{q_1+q_2-q_1q_2}-\eta_1=0.
\end{align}
Rearranging and using complementary slackness \cite{boyd} directly gives (\ref{eq_w1_opt}). Similar arguments yield (\ref{eq_w2_opt}).
\end{Proof}

Lemma~\ref{thm_waiting} indicates that the optimal waiting policy has a {\it threshold} structure. Basically, the optimal waiting function $w^*(\cdot)$ is given by
\begin{align}
w^*(y)=\left[\lambda-\mathbb{E}\left[X\right]-y\right]^+,
\end{align}
with $y$ being the realizing of the starting AoI $Y$. This means that the transmitter does not send a new measurement unless the AoI grows above the threshold $\lambda-\mathbb{E}\left[X\right]$. To have an operational significance, however, the value of such threshold needs to be positive. Otherwise, waiting is never optimal. In the next lemma we show that this is indeed the case at the optimal $\lambda^*$. The proof is in Appendix~\ref{apndx_thm_lmda_lb}.

\begin{lemma} \label{thm_lmda_lb}
$\lambda^*$ that solves $p(\lambda^*)=0$ is such that $\lambda^*>\mathbb{E}\left[X\right]$.
\end{lemma}

We now have three regions in which $\lambda^*$ can lie:
\begin{align}
\mathcal{R}_1\triangleq&\left\{\lambda:~\mathbb{E}\left[X\right]<\lambda\leq\mathbb{E}\left[X\right]+n\right\}; \\
\mathcal{R}_2\triangleq&\left\{\lambda:~\mathbb{E}\left[X\right]+n<\lambda\leq\mathbb{E}\left[X\right]+n+m\right\}; \\
\text{and }\mathcal{R}_3\triangleq&\left\{\lambda:~\mathbb{E}\left[X\right]+n+m<\lambda\right\}.
\end{align}
For $\lambda^*\in\mathcal{R}_1$, the zero-wait policy is optimal, i.e., $w_1^*=w_2^*=0$. While for $\lambda^*\in\mathcal{R}_2$, waiting is only optimal following a successful transmission from the first decoding attempt, i.e., $w_1^*>0$ and $w_2^*=0$. Finally for $\lambda^*\in\mathcal{R}_3$, waiting is always optimal following a successful transmission, i.e., $w_1^*>0$ and $w_2^*>0$. In the next theorem, we derive necessary and sufficient conditions for the optimal $\lambda^*$ to lie in each region. The proof is in Appendix~\ref{apndx_thm_opt_sol}.

\begin{theorem} \label{thm_opt_sol}
$\lambda^*$ that solves $p(\lambda^*)=0$ is given by
\begin{align}
\lambda^*=
\begin{cases}
\overline{\lambda}\in\mathcal{R}_1,\quad&\text{if }n\geq m\sqrt{1-q_1} \\
\underline{\lambda}\in\mathcal{R}_2,\quad&\text{otherwise}
\end{cases},
\end{align}
where
\begin{align}
\overline{\lambda}&\triangleq\mathbb{E}\left[Y\right]+\frac{\frac{1}{2}\mathbb{E}\left[X^2\right]}{\mathbb{E}\left[X\right]}, \label{eq_ovr_lmda} \\
\underline{\lambda}&\triangleq \mathbb{E}\left[X\right]+n+\frac{\sqrt{\left(\mathbb{E}\left[X\right]\right)^2-\frac{2q_1}{q_1+q_2-q_1q_2}C_{XY}}-\mathbb{E}\left[X\right]}{\frac{q_1}{q_1+q_2-q_1q_2}}, \label{eq_undr_lmda}
\end{align}
with $C_{XY}\triangleq\left(\mathbb{E}\left[X\right]\right)^2+n\mathbb{E}\left[X\right]-\frac{1}{2}\mathbb{E}\left[X^2\right]-\mathbb{E}\left[Y\right]\mathbb{E}\left[X\right].$
\end{theorem}

Theorem~\ref{thm_opt_sol} indicates that the optimal waiting policy is such that one can either wait following successful decoding from the first attempt, or do not wait at all. It is therefore {\it not} optimal to wait following successful decoding from the second attempt, since this renders the AoI relatively high and incentivizes sending a new packet right away. Clearly, choosing the best policy now depends on the choice of $n$ and $m$, which governs the channel behavior through $q_1$ and $q_2$. We discuss this in the next section, along with some examples.

\section{How to Choose $n$ and $m$: Examples}

In this section, we discuss how the choices of the codeword length $n$ and IR length $m$ impact the AoI. Let us denote the optimal solution of problem (\ref{opt_sd}) by $\rho^*$. In view of Theorem~\ref{thm_opt_sol}, let us define the following quantities:
\begin{align}
\overline{\lambda}^*\triangleq&\min_{n,m:~n\geq m\sqrt{1-q_1}}\quad\overline{\lambda},\\
\underline{\lambda}^*\triangleq&\min_{n,m:~n< m\sqrt{1-q_1}}\quad\underline{\lambda},
\end{align}
where $\overline{\lambda}$ and $\underline{\lambda}$ are given by (\ref{eq_ovr_lmda}) and (\ref{eq_undr_lmda}), respectively. Therefore, it is direct to see that
\begin{align}
\rho^*=\min\left\{\overline{\lambda}^*,\underline{\lambda}^*\right\}.
\end{align}

Observe that computing the exact values of $\overline{\lambda}^*$ and $\underline{\lambda}^*$ involves solving non-linear integer programs. One approach to alleviate the difficulty of such step is to relax the integer constraints on $n$ and $m$, i.e., solve for $n,m\in\mathbb{R}_+$, and then project the solution onto the feasible set in $\mathbb{Z}_+$. Note that solving for $\rho^*$ can be done offline before communication. Our goal here though is to show how $n$ and $m$ impact the waiting policy and the AoI. Thus, in what follows we follow a grid search approach to characterize $\rho^*$ numerically. 

\begin{figure}[t]
\center
\includegraphics[scale=.35]{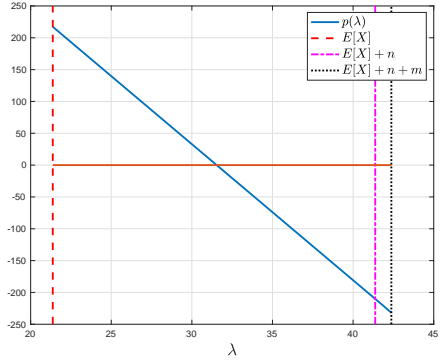}
\caption{$\ell=15$, $n=20$ and $\epsilon=0.1$. The optimal IR length is $m=1$.}
\label{fig_p_lambda_good_channel}
\vspace{-.2in}
\end{figure}

\begin{figure}[t]
\center
\includegraphics[scale=.35]{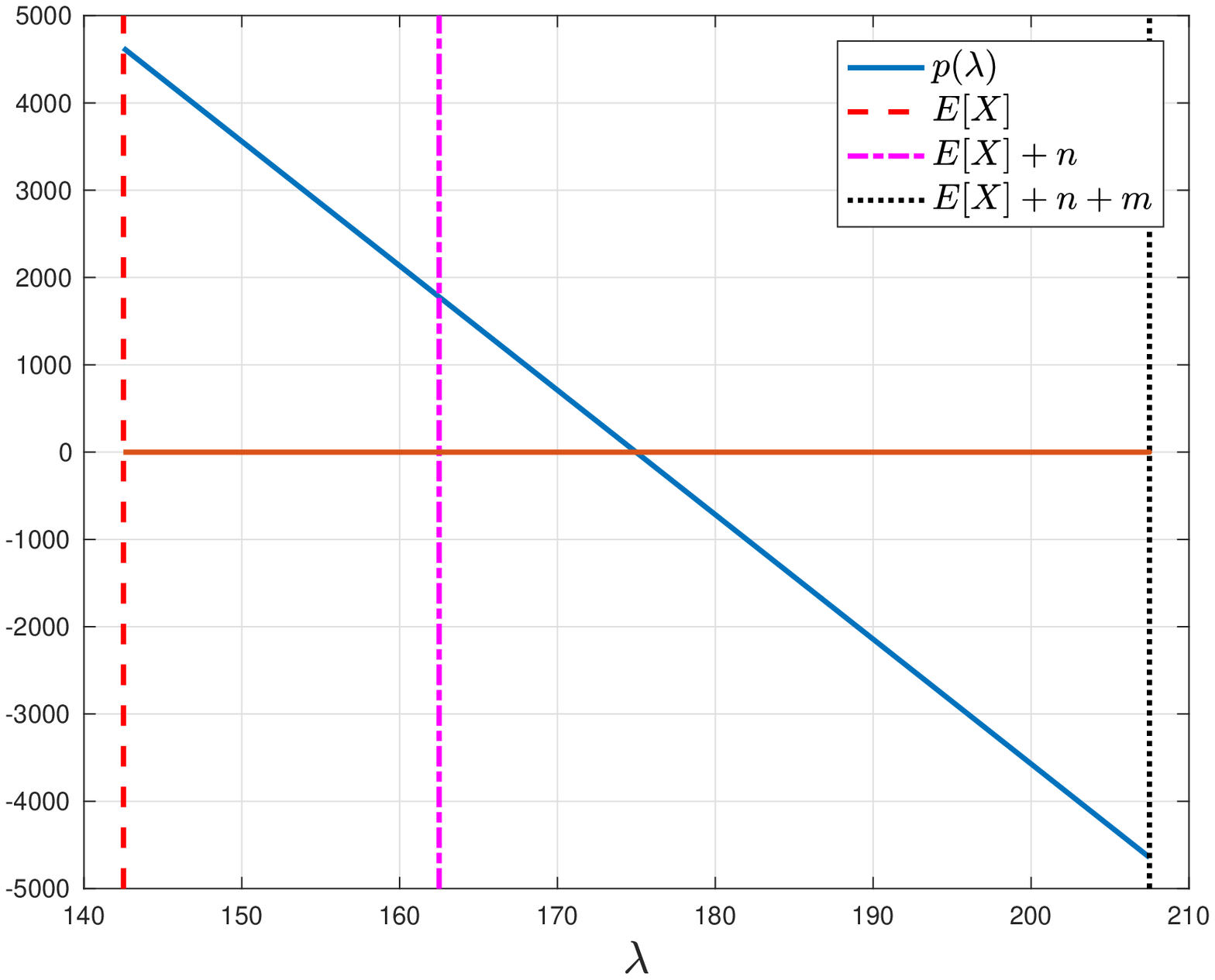}
\caption{$\ell=15$, $n=20$ and $\epsilon=0.4$. The optimal IR length is $m=45$.}
\label{fig_p_lambda_bad_channel}
\vspace{-.1in}
\end{figure}

\begin{figure}[t]
\center
\includegraphics[scale=.35]{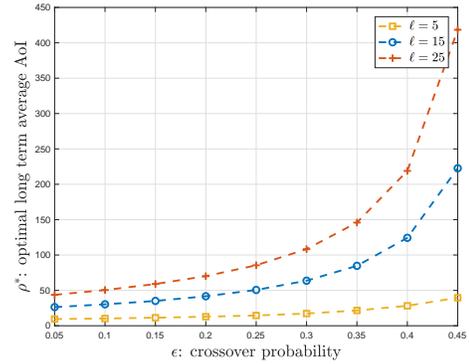}
\caption{Optimal AoI versus cross over probability for different values of $\ell$.}
\label{fig_aoi_vs_eps}
\vspace{-.2in}
\end{figure}

We consider the communication channel to be a binary symmetric channel (BSC) with crossover probability $\epsilon\in\left(0,\frac{1}{2}\right)$. We use an $\left(n+m,\ell\right)$ MDS code, from which we take a punctured $\left(n,\ell\right)$ code by removing $m$ columns from its generator matrix. Note that $\left(n,\ell\right)$, for $n\geq\ell$, is also MDS \cite{feyling-punctured-mds-codes}. We use the punctured code for the first transmission attempt, and then append the IR bits from the original code in the second transmission attempt if needed. This allows us to write $q_1=\sum_{l=1}^{\lfloor{\frac{n-\ell}{2}}\rfloor}\binom{n}{l}\epsilon^l\left(1-\epsilon\right)^{n-l}$, and $q_2=\sum_{l=1}^{\lfloor{\frac{n+m-\ell}{2}}\rfloor}\binom{n+m}{l}\epsilon^l\left(1-\epsilon\right)^{n+m-l}$, where $\lfloor x\rfloor$ denotes the highest integer not larger than $x$. We show how the optimal $m$ behaves as $\epsilon$ varies for fixed $\ell=15$ bits and $n=20$ bits. For $\epsilon=0.1$, i.e., when the channel is relatively good, the optimal $m=1$ bit, and the optimal AoI $\lambda^*\approx31.54$ time units. It is seen from Fig.~\ref{fig_p_lambda_good_channel} that $\lambda^*\in\mathcal{R}_1$ in this case, and zero-waiting is optimal. On the other hand, for $\epsilon=0.4$, i.e., when the channel is relatively bad, the optimal $m=45$ bits, and the optimal AoI $\lambda^*\approx174.97$ time units. From Fig.~\ref{fig_p_lambda_bad_channel}, we see that $\lambda^*\in\mathcal{R}_2$ and $w_1^*>0$ in this case. Such results show that the optimal choice of the IR length and the waiting policy that reduce the AoI varies according to the channel conditions. In Fig.~\ref{fig_aoi_vs_eps} we plot $\rho^*$ versus $\epsilon$ for different values of $\ell$. Here, we optimize both $n$ and $m$. We see from the figure that the age grows with the crossover probability, with an increasing growth rate as $\epsilon$ approaches $0.5$, the value at which no information can be conveyed through the BSC. We also see that the growth rate is more intense for larger values of $\ell$. Note, however, that choosing $\ell$ may depend on some sensor aspects, such as the sampling frequency, and also on the amount of distortion that can be tolerated in the system when converting raw measurements to $\ell$-bit messages.

\section{Conclusion}

Designing timely channel coding schemes has been examined for a transmitter-receiver pair communicating through a noisy channel. A HARQ protocol has been employed in which IR bits are transmitted to enhance probability of successful decoding. An idle waiting period has been introduced following successful transmissions. The age-minimal waiting policy has been shown to have a threshold structure that depends on the codeword and IR lengths. The optimal threshold has been obtained analytically in closed-form. A discussion on how to choose the best codeword and IR lengths for a binary symmetric channel with MDS codes has been presented.

\appendix

\subsection{Proof of Lemma~\ref{thm_lmda_lb}} \label{apndx_thm_lmda_lb}

We show this by contradiction. Assume that $\lambda^*\leq\mathbb{E}\left[X\right]$. By (\ref{eq_w1_opt}) and (\ref{eq_w2_opt}), this means that $w_1^*=w_2^*=0$. Substituting this in (\ref{eq_exp_L}) and (\ref{eq_exp_Q}), this further means that
\begin{align}
\lambda^*=\frac{\mathbb{E}\left[Q\right]}{\mathbb{E}\left[L\right]}=\mathbb{E}\left[Y\right]+\frac{\frac{1}{2}\mathbb{E}\left[X^2\right]}{\mathbb{E}\left[X\right]}. \label{eq_lmda_zero_wait}
\end{align}
Now $\lambda^*\leq\mathbb{E}\left[X\right]$ implies
\begin{align}
\mathbb{E}\left[Y\right]\mathbb{E}\left[X\right]+\frac{1}{2}\mathbb{E}\left[X^2\right]\leq\left(\mathbb{E}\left[X\right]\right)^2.
\end{align}
Substituting (\ref{eq_exp_Y}), (\ref{eq_exp_X}) and (\ref{eq_exp_X2}) in the above and multiplying both sides by $\left(q_1+q_2-q_1q_2\right)^2$ we get that
\begin{align}
&\left(n\left(q_1+q_2-q_1q_2\right)+m\left(1-q_1\right)q_2\right)\left(n+m\left(1-q_1\right)\right) \nonumber \\
&+\frac{1}{2}\left(n+m\right)^2\left(2-q_1-q_2+q_1q_2\right)-m\left(n+m\right)q_1 \nonumber \\
&+\frac{1}{2}m^2q_1\left(q_1+q_2-q_1q_2\right)\leq\left(n+m(1-q_1)\right)^2,
\end{align}
which is equivalent to having
\begin{align}
&\frac{1}{2}\left(n+m\right)^2\left(2-q_1-q_2+q_1q_2\right)-m\left(n+m\right)q_1 \nonumber \\
&+\frac{1}{2}m^2q_1\left(q_1+q_2-q_1q_2\right)\leq\left(n+m\right)^2(1-q_1)(1-q_2) \nonumber \\
&-m(n+m)q_1(1-q_1)(1-q_2).
\end{align}
One final rearrangement of the above, followed by dividing both sides by $q_1+q_2-q_1q_2$, gives
\begin{align}
\frac{1}{2}\left(n+m\right)^2+\frac{1}{2}m^2q_1&\leq m\left(n+m\right)q_1, \\
\iff\frac{1}{2}n^2+mn(1-q_1)+\frac{1}{2}m^2&\leq\frac{1}{2}m^2q_1,
\end{align}
which cannot be true, indicating a contradiction.

\subsection{Proof of Theorem~\ref{thm_opt_sol}} \label{apndx_thm_opt_sol}

Observe that since $p(\lambda)$ is monotonically decreasing, having $\lambda^*\in\mathcal{R}_1$ is equivalent to having
\begin{align}
p\left(\mathbb{E}\left[X\right]+n\right)\leq0.
\end{align}
Note that for $\lambda=\mathbb{E}\left[X\right]+n$, $w_1=w_2=0$. Substituting this in (\ref{eq_exp_L}) and (\ref{eq_exp_Q}), the above inequality condition is equivalent to
\begin{align} \label{eq_R_1_condition}
\mathbb{E}\left[Y\right]\mathbb{E}\left[X\right]+\frac{1}{2}\mathbb{E}\left[X^2\right]\leq\left(\mathbb{E}\left[X\right]\right)^2+n\mathbb{E}\left[X\right].
\end{align}
Multiplying both sides of (\ref{eq_R_1_condition}) by $\left(q_1+q_2-q_1q_2\right)^2$ and proceeding via similar simplifications as those in the proof of Lemma~\ref{thm_lmda_lb}, one can show that (\ref{eq_R_1_condition}) is equivalent to
\begin{align}
\frac{1}{2}\left(n+m\right)^2+\frac{1}{2}m^2q_1&\leq m\left(n+m\right)q_1 + n\left(n+m(1-q_1)\right), \\
\iff m^2\left(1-q_1\right)&\leq n^2.
\end{align}
Therefore, $\lambda^*\in\mathcal{R}_1\iff n\geq m\sqrt{1-q_1}$. Finally, it is direct to see that such $\lambda^*$ is given by $\overline{\lambda}$ of (\ref{eq_ovr_lmda}) in this case, and that, by (\ref{eq_R_1_condition}), $\overline{\lambda}$ indeed lies in $\mathcal{R}_1$.

Now let us assume that $n<m\sqrt{1-q_1}$. From the above, this is equivalent to having $p\left(\mathbb{E}\left[X\right]+n\right)>0$. Therefore, having $\lambda^*\in\mathcal{R}_2$ in this case is equivalent to having
\begin{align}
p\left(\mathbb{E}\left[X\right]+n+m\right)\leq0.
\end{align}
Note that for $\lambda=\mathbb{E}\left[X\right]+n+m$, $w_1=m$ and $w_2=0$. Substituting this in (\ref{eq_exp_L}) and (\ref{eq_exp_Q}), the above inequality becomes
\begin{align} \label{eq_R_2_aux}
&\frac{q_1mn}{q_1+q_2-q_1q_2}+\mathbb{E}\left[Y\right]\mathbb{E}\left[X\right]+\frac{1}{2}\mathbb{E}\left[X^2\right] \nonumber \\
&+\mathbb{E}\left[X\right]\frac{q_1m}{q_1+q_2-q_1q_2}+\frac{1}{2}\frac{q_1m^2}{q_1+q_2-q_1q_2} \nonumber \\
&\hspace{.2in}\leq \left(\mathbb{E}\left[X\right]+n+m\right)\left(\mathbb{E}\left[X\right]+\frac{q_1m}{q_1+q_2-q_1q_2}\right).
\end{align}
This can be further simplified into the following:
\begin{align} \label{eq_R_2_condition}
\mathbb{E}\left[Y\right]\mathbb{E}\left[X\right]+&\frac{1}{2}\mathbb{E}\left[X^2\right]\leq\left(\mathbb{E}\left[X\right]\right)^2+n\mathbb{E}\left[X\right] \nonumber \\
&+m\mathbb{E}\left[X\right]+\frac{1}{2}\frac{m^2q_1}{q_1+q_2-q_1q_2}.
\end{align}
One can clearly notice the resemblance between (\ref{eq_R_1_condition}) and (\ref{eq_R_2_condition}). Proceeding as done after (\ref{eq_R_1_condition}), (\ref{eq_R_2_condition}) is equivalent to
\begin{align}
m^2(1-q_1)\leq n^2+m\left(n+m(1-q_1)\right)+\frac{1}{2}m^2q_1,
\end{align}
which is always satisfied, upon cancelling $m^2(1-q_1)$ from both sides. This shows that $\lambda^*\in\mathcal{R}_2 \iff n<m\sqrt{1-q_1}$.

It finally remains to derive the value of $\underline{\lambda}$ of (\ref{eq_undr_lmda}). To do so, we solve $p(\lambda^*)=0$ under the condition that $\lambda^*\in\mathcal{R}_2$. Equivalently, we replace $m$ in (\ref{eq_R_2_aux}) by $w_1$; solve for $w_1^*$ that satisfies (\ref{eq_R_2_aux}) with equality; and then substitute $\lambda^*=w_1^*+\mathbb{E}\left[X\right]+n$. Toward that, we solve
\begin{align}
\frac{1}{2}\frac{q_1}{q_1+q_2-q_1q_2}\left(w_1^*\right)^2+\mathbb{E}\left[X\right]w_1^*+C_{XY}=0,
\end{align}
with $C_{XY}$ as defined in the theorem. Solving the above and adding $\mathbb{E}\left[X\right]+n$ directly gives $\underline{\lambda}$ of (\ref{eq_undr_lmda}). Finally, since $n<m\sqrt{1-q_1}$, therefore (\ref{eq_R_1_condition}) does not hold and $C_{XY}$ is strictly negative. This shows that $w_1^*>0$ and that $\underline{\lambda}$ indeed lies in $\mathcal{R}_2$. This completes the proof of the theorem.


\end{document}